\documentclass[pre,a4paper]{revtex4}
\usepackage{graphicx}
\usepackage{subfigure}
\usepackage{latexsym}

\begin{document}

\author{Marco Cosentino Lagomarsino}

\affiliation{FOM Institute for Atomic and Molecular Physics (AMOLF),
  Kruislaan 407, 1098 SJ Amsterdam, The Netherlands}

\email{cosentino-lagomarsino@amolf.nl, dogterom@amolf.nl.}

\author{Marjolein Dijkstra} 

\affiliation{Debye Institute, Soft
  Condensed Matter Physics, Utrecht University, Princetonplein 5, 3584
  CC Utrecht, The Netherlands} 

\email{m.dijkstra@phys.uu.nl.}

\author{ Marileen Dogterom}
\affiliation{FOM Institute for Atomic and Molecular Physics (AMOLF),
  Kruislaan 407, 1098 SJ Amsterdam, The Netherlands}

\title{Isotropic-Nematic transition of long thin hard spherocylinders
  confined in a quasi-two-dimensional planar geometry.}


\begin{abstract}
  We present computer simulations of long thin hard spherocylinders in
  a narrow planar slit. We observe a transition from the isotropic to
  a nematic phase with quasi-long-range orientational order upon
  increasing the density.  This phase transition is intrinsically two
  dimensional and of the Kosterlitz-Thouless type. The effective
  two-dimensional density at which this transition occurs increases
  with plate separation.  We qualitatively compare some of our results
  with experiments where microtubules are confined in a thin slit,
  which gave the original inspiration for this work.
\end{abstract}

\maketitle

\section{Introduction}

Two-dimensional systems behave generically in a qualitatively distinct
way from three-dimensional ones. In experiments, they are often not
strictly confined to a mathematical surface, but also span a small
region in the transverse dimension. A situation where two hard walls
very close to each other confine the system occurs between the two
extremes of bulk and two-dimensions, and, for this reason, is
interesting to analyze.
We study the phase behavior of a system confined between two parallel
planar plates separated by small distances as a function of the plate
separation and relate it to the corresponding strictly two-dimensional
system.  We address this question, using computer simulations, for the
Isotropic-Nematic (I-N) transition of thin, hard spherocylinders.
The inspiration for this problem comes from a biological system,
cortical microtubules in plant cells, and by in vitro experiments on
microtubules confined in thin samples.  Microtubules are tubular
polymers of the protein tubulin with an aspect ratio of a few
hundreds. They are the stiffest polymers available to eukaryotic
cells, with a persistence length of about 3 millimeters~\cite{marcel},
two orders of magnitude greater than their typical length. With
microtubules one can construct a good model system for the
quasi-two-dimensional regime of (broadly polydisperse) hard rods. In
this paper, we focus on the results of our simulations, while also
commenting on the possible links to the experimental system.

It is known~\cite{frenkel,bates} that a system of hard needles or high
(\(>9\)) aspect ratio disco-rectangles in two dimensions exhibits,
similarly to the xy-model, a nematic transition to quasi-long-range
order of the Kosterlitz-Thouless kind~\cite{koth}. The role of
disclinations in this system has been investigated for different
two-dimensional geometries~\cite{schmidt_def}.  A related problem is
that of surface effects in presence od a single
wall~\cite{schmidt_sph,poniewirski,MarjoleinEPL}.  Several theoretical
and simulation studies were focussed on fluids of hard rods in contact
with hard walls. Chen and Cui performed density functional theory
calculations for a fluid of hard semiflexible polymers near a hard
wall. They show a weakly first-order uniaxial to biaxial transition
(signature of two dimensional ordering in the plane defined by the
wall) at a bulk density far below that of the bulk I-N transition
\cite{chen}. They also observe that the formation of a biaxial nematic
film at the wall-isotropic fluid interface with the director parallel
to the wall. The thickness of this film appears to diverge as bulk I-N
coexistence is approached \cite{chen}. A surface induced continuous
transition from a uniaxial to a biaxial phase prior to complete
wetting was also predicted by density-functional theory of the Zwanzig
model \cite{MarjoleinEPL,ReneJCP}, in which the orientations of the
particles are restricted to three orthogonal directions, and confirmed
by simulations for freely rotating
spherocylinders~\cite{MarjoleinEPL,MarjoleinPRE}. Returning to the
case of a fluid confined by two parallel hard walls, it was found by
simulations of freely rotating spherocylinders~\cite{MarjoleinPRE} and
by density functional theory calculations of the Zwanzig
model~\cite{ReneJCP} that the surface-induced uniaxial to biaxial
transition is prior to a first-order capillary nematization transition
at larger bulk densities, which terminates in a capillary critical
point when the wall separation is about twice the length of the rods.

This paper is organized as follows.  In section \ref{sec:model} we
present our model of hard spherocylinders confined between two
parallel planar hard walls, and we introduce the quantities that are
measured, namely, the nematic order parameter and the orientational
correlation function. In section \ref{sec:op}, we present the results
and we discuss the polydisperse case and the possible connections with
the experimental system.  Our main result is that similarly to the
strictly two-dimensional system, there is no true transition to long
range order in the thermodynamic limit.  A thickness-dependent
transition to quasi-long range orientational order of the Kosterlitz
and Thouless kind is present instead.

This is in qualitative agreement with what is observed experimentally
in a system of microtubules.

\section{A fluid of hard spherocylinders confined in a
quasi-two-dimensional planar slit} \label{sec:model}

\begin{figure}[!ht]
  \centering
  \includegraphics[scale=.7]{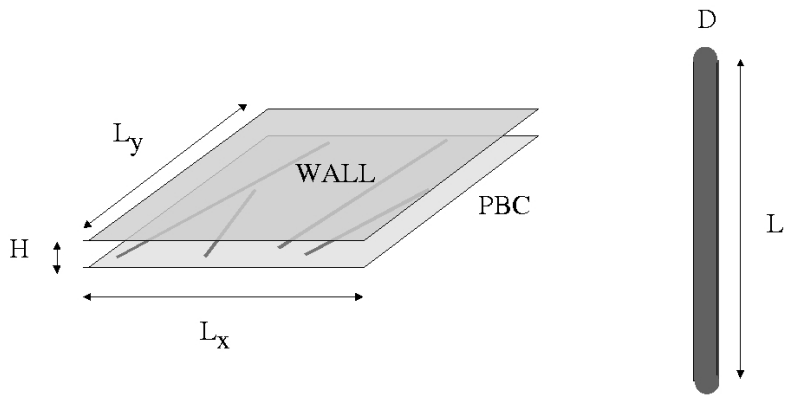}
  \caption{A schematic picture of the system, which consists of a slit
    of thickness $h=H/D$, enclosed by two planar hard walls in the
    $x$-$y$ plane with dimensions $L_x \times L_y$. A particle
    (spherocylinder) is described by a cylinder of length $L$ and
    diameter $D$ capped by two hemispheres of the same
    diameter. Periodic boundary conditions are applied laterally. }
  \label{introfig}
\end{figure}

We perform Monte-Carlo simulations for a fluid of hard spherocylinders
with a (mean) length-to-diameter ratio  $L/D=320$, corresponding to
microtubules of 8 \(\mu m\) in length and 25 $nm$ in diameter,
confined between two planar hard walls with an area of $L_x \times
L_y$ in the $x-y$ plane at distance $h=H/D$ in the $z$ direction (see
Fig. \ref{introfig}). Periodic boundary conditions are employed in the
$x-$ and $y-$direction. The slit width \(h \) varies from 1 to 120. We
perform simulations both of a system of monodisperse rods and a system
of polydisperse rods with an exponential length distribution, which
resembles closely the experiment with microtubules (see Section
\ref{sec:exp}).  The simulations consist of about \(2\cdot10^6\) Monte
Carlo sweeps, where one sweep equals one attempted move per particle.
Typically more than \( 10^5\) Monte Carlo sweeps were allowed for
equilibration. The number of particles $N$ in a simulation is up to
5000, although typically lower than 2000.

We perform simulations for monodisperse spherocylinders for varying
dimensionless densities:
\begin{displaymath}
  C = \frac{N}{V} (L+D)^2 D,
\end{displaymath}
where $N$ is the number of particles, $V=L_x L_y H$ is the total
volume. We measure the eigenvalues of the standard \(3\times 3\)
nematic order-parameter tensor
\begin{displaymath}
  Q_{\alpha \beta} = \left\langle \frac{1}{N} \sum_{i=1}^N \left(
  \frac{3}{2} u_{\alpha}^i u_{\beta}^i - \frac{\delta_{\alpha
  \beta}}{2} \right)     \right\rangle,
\end{displaymath}
where \(\alpha, \beta = x,y,z\); \( u_{\alpha}^i\) is the \(\alpha
\) component of the unit vector defining the orientation of
particle $i$, and \(\delta_{\alpha \beta} \) is Kronecker's delta.
Diagonalizing \( Q_{\alpha \beta} \) gives the orientational order
parameters.  As the typical cylindrical symmetry of the nematic
state is broken a priori by the geometric constraints on the
system, the most suitable order parameter \(\Delta\) is
proportional to the difference between the two highest eigenvalues
of the average nematic tensor, normalized with a factor of 2/3 to
make it lie in the interval \([0,1]\).  It is easy to realize that
when \( h= 1 \) this quantity reduces
to the two-dimensional nematic order parameter \cite{frenkel} and for
larger $h$ it can be identified with the biaxial order
parameter~\cite{MarjoleinEPL}. In fact, for \(h=1\), one of the three
eigenvalues of Q will be \(-1/2\) due to the geometric confinement, so that
the submatrix defined by the other two eigenvalues will coincide with
the exception of a constant factor with the two-dimensional nematic
order parameter \(2\times 2\) tensor, and
\begin{displaymath}
  \Delta \sim S = \left\langle \frac{1}{N} \sum_{i=1}^{N}
  cos(2\theta_i) \right\rangle,
\end{displaymath}
where \(\theta_i\) is the planar smallest angle formed by the
\emph{i}-th particle with the nematic director, and S is the
2-dimensional orientational order parameter.  We also examine the
density and orientational order parameter \(\Delta \) profiles along
the axis orthogonal to the walls. These are measured using bins with a
typical width of about \(D\), accumulated over the sweeps in the
simulation \cite{MarjoleinEPL}. Lastly, we measure the orientational
correlation function
\begin{displaymath}
  g_2(r) = \langle(cos(\theta(0) - \theta(r))\rangle,
\end{displaymath}
which describes the decay of the long wavelength fluctuations in the
orientation.

\section{Results}

We will first discuss the monodisperse case. The polydisperse system
will be examined briefly, motivated by its relevance to experiments
with microtubules.

\subsection*{Order parameter and size effects.}
\label{sec:op}

\begin{figure}[!htb]
  \centering
  \includegraphics[scale=.7]{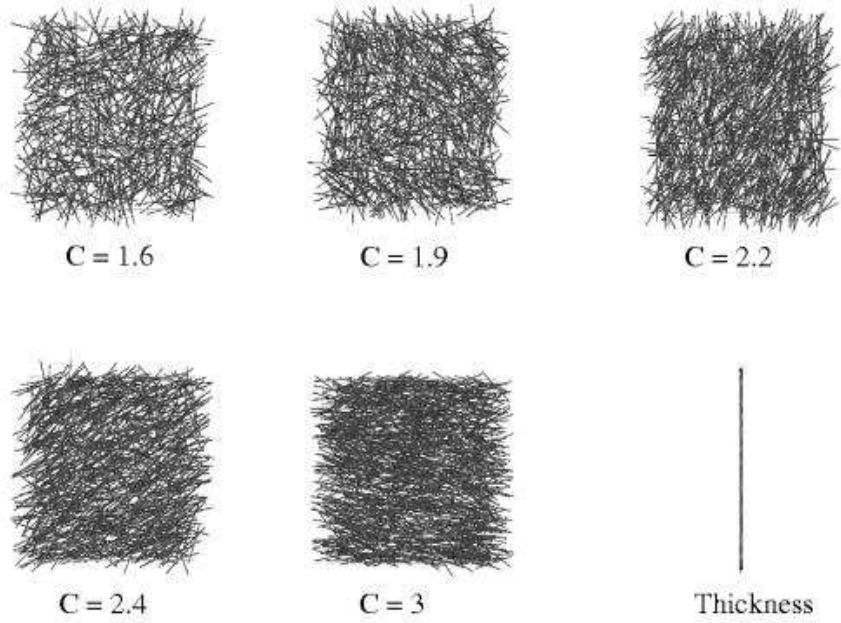}
  \caption{Snapshots of the configuration of the system projected on
    the $xy$ plane for different values of the dimensionless density
    $C=N/V (L+D)^2 D$. Here \(h = 20\), \(L_{xy} = 1400 D\), $N$
    varies from 600 to 1150 and the equilibration time is \( 10^6\)
    sweeps. The diameters of the rods are not drawn to scale.}
  \label{fig:snapshots}
\end{figure}

\begin{figure}[!htb]
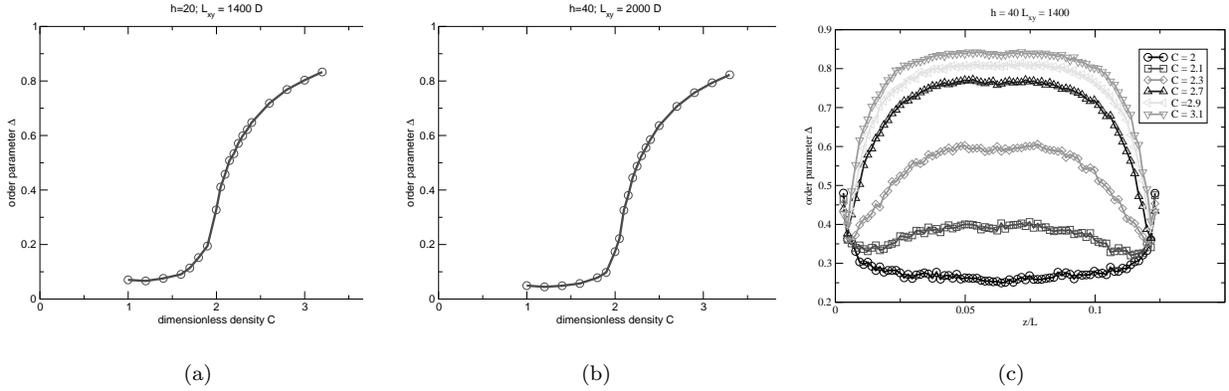

  \centering
  \subfigure[]{\includegraphics[scale=.24]{transh20.eps}}
  \subfigure[]{\includegraphics[scale=.24]{transh40.eps}}
  \subfigure[]{\includegraphics[scale=.24]{op_profile.eps}}
  \caption{(a) and (b).  Order parameter \(\Delta\) as a function of
    non-dimensional density for \(L_{xy} = 1400 D\). (a) \(h = 20\)
    (b) \(h = 40 \). (c) $z$ profile of the order parameter for \(h =
    40\), \(L_{xy} = 1400 D\) and increasing density. The transition
    density \(C_{\textrm{trans}}\) can be defined through the
    intersection of the tangents constructed on the low density part
    and the linearly increasing part of graphs like (a) and (b). }
  \label{fig:OPD}
\end{figure}

Fixing the lateral size $L_x=L_y=L_{xy}$  of the system, we
measure the order parameter \(\Delta \) as a function of $C$. We
find a transition to an ordered state for a density which depends
slightly  on $h$. In figure \ref{fig:snapshots} we show typical
snapshots of the fluid of spherocylinders for different densities
of rods and a lateral view of the simulation box to illustrate its
thickness. A transition from a uniaxial phase (in which the rods
are randomly oriented in the $xy$ plane)  to a biaxial phase (the
rods have a preferred orientation in the $xy$ plane) is apparent
from visual analysis of these pictures.
The order parameter and its profile along the $z$-axis as a
function of density are plotted for a few examples in figure
\ref{fig:OPD}. We define arbitrarily the apparent transition density
\(C_{\textrm{trans}}\) as the density at the intersection of the
tangents constructed on the low density part and the linearly
increasing part of \( \Delta(C) \) at fixed density (see the caption
of figure \ref{fig:OPD}).

\begin{figure}[!ht]
  \centering
  \includegraphics[scale=.3]{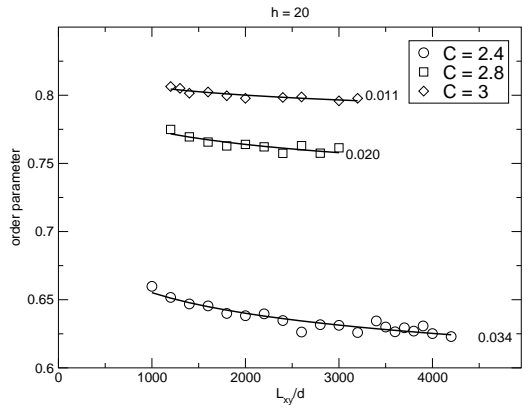}
  \caption{Finite size effects in the measurement of the order
    parameter $\Delta$. The graph shows the decay of $\Delta$ for
    three dimensionless densities as a function of the lateral size
    \(L_{xy}\) when the vertical size \(h =20\) is kept fixed. The
    numbers to the right of the curves are the exponent from power law
    fits. The decay of the order parameter for increasing size causes
    a shift in the apparent transition.}
  \label{fig:siz}
\end{figure}

The apparent phase transition from a uniaxial to biaxial phase,
however, depends in all instances on the lateral size of the system.
In fact, the order parameter drops with increasing size at fixed $C$,
in a way that the apparent transition density increases. We measured
this effect consistently for different transverse sizes (\(h =
5,20,40,60\)). In figure \ref{fig:siz} we report an example for \(h=
20\). 

This fact gives a hint that, as in the strictly two-dimensional
system, there could be no true phase transition in the thermodynamic
limit, but a transition to quasi-long-range order.  To establish this,
it is necessary to test whether the orientational correlation function
shows an algebraic decay (see below).

\begin{figure}[!ht]
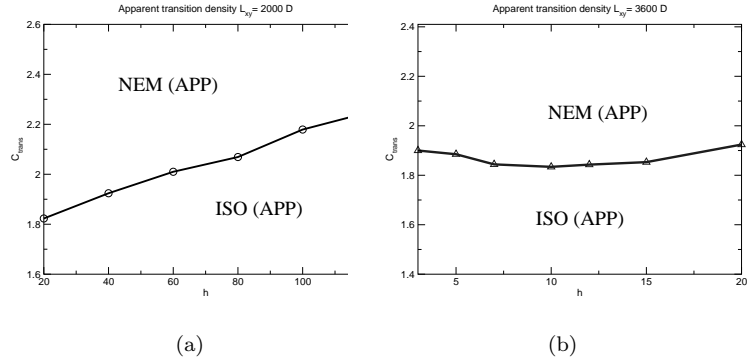

  \centering
     \subfigure[]{\includegraphics[scale=.22]{app_hdep2.eps}}
     \subfigure[]{\includegraphics[scale=.22]{app_hdep1.eps}}
  \caption{Phase diagrams for the apparent nematic transition at fixed
    lateral size $L_{xy}$.  The reduced uniaxial-biaxial transition
    density \(C_{\textrm{trans}}\) for different values of \(h\).  (a)
    Lateral size \(L_{xy} = 2000 D\), thickness from \(h = 20\) up to
    \(h = 120\); the number of particles $N$ for these runs is up to
    7000. (b) Lateral size \(L_{xy} = 3600 D\), thickness up to
    \(h=20\), $N$ up to 6000. The two critical lines in (a) and (b) do
    not connect continuously at \(h = 20\) because of the finite size
    effects.}
  \label{fig:hdep}
\end{figure}

For comparison with real finite size systems it is interesting to look
at the apparent transition density \(C_{\textrm{trans}}\), at fixed
lateral size $L_{xy}$, varying the small transverse size $h$. Two
examples of the resulting phase diagrams are reported in figure
\ref{fig:hdep}. The transition density is roughly constant for
thicknesses lower than \(h=15\) (see Fig. \ref{fig:hdep}b) and
increases almost linearly for thicker boxes (see Fig.
\ref{fig:hdep}a).

\subsection*{Correlation function}
\label{sec:g2} 

\begin{figure}[!h]
  \centering
  \includegraphics[scale=.3]{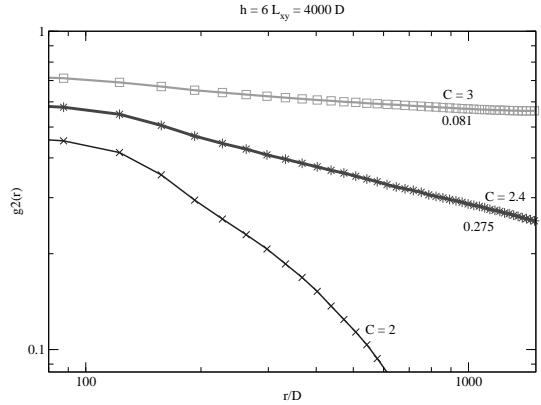}
  \caption{Log-log plot of the orientational correlation function
    \(g_2(r)\) for different values of the density, in the case
    \(h=6\). For low density this quantity has exponential decay,
    which becomes algebraic with increasing $C$. The numbers next to
    the curves indicate the decay exponents from power-law fits.  By
    convention, the transition is defined to be where the exponent of
    the correlation function crosses 1/4, in this instance around \( C
    = 2.4 \).}
  \label{fig:g2ex}
\end{figure}

As we argued in the previous section, because of the finite size
effects (decrease of \(\Delta\) with increasing size), little can be
concluded about the nature of the phase transition, or even its
existence based on the measurements of the order parameter.  These
finite size effects, together with the presence of a phase transition
to quasi-long-range order in the strictly two-dimensional system, make
it sensible to investigate the decay of long wavelength orientational
fluctuations. This is a delicate matter due to the very long
relaxation times of these soft modes. We investigated the transition
for integer values of \(h\) up to \(h=10\), with a simulation box
having a lateral size \( L_{xy} = 4000D\). Above this thickness, the
number of particles was too high to achieve relaxation in reasonable
times. In all these cases \(g_2(r)\) showed a transition from
exponential to power law decay with increasing density. One example,
for \(h = 6 \), is plotted in figure \ref{fig:g2ex}.  This transition
does not depend on the size of the system, which just needs to be
large enough so that the correlation of the modes is not affected by
the periodic lateral boundary conditions.

\begin{figure}[!h]
  \centering
  \includegraphics[scale=.3]{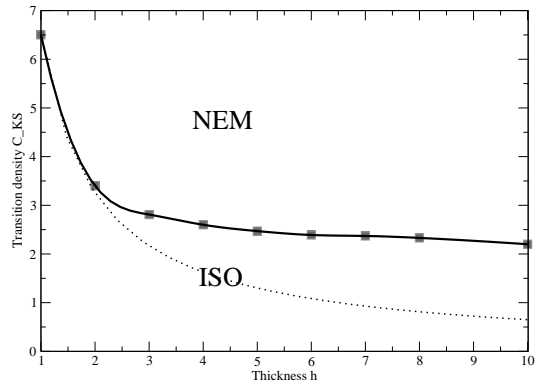}
  \caption{Phase diagram for the nematic transition in quasi two
    dimensional geometry. The Kosterlitz-Thouless transition density
    \(C_{\textrm{trans}}^{KS} \) is plotted as a function of the
    system thickness \(h\). Its decay is slower than the naively
    expected \(1/h\) indicated by the dotted line.}
  \label{fig:KShdep}
\end{figure}

\begin{figure}[!htbp]
  \centering
  \subfigure[]{\includegraphics[scale=.45]{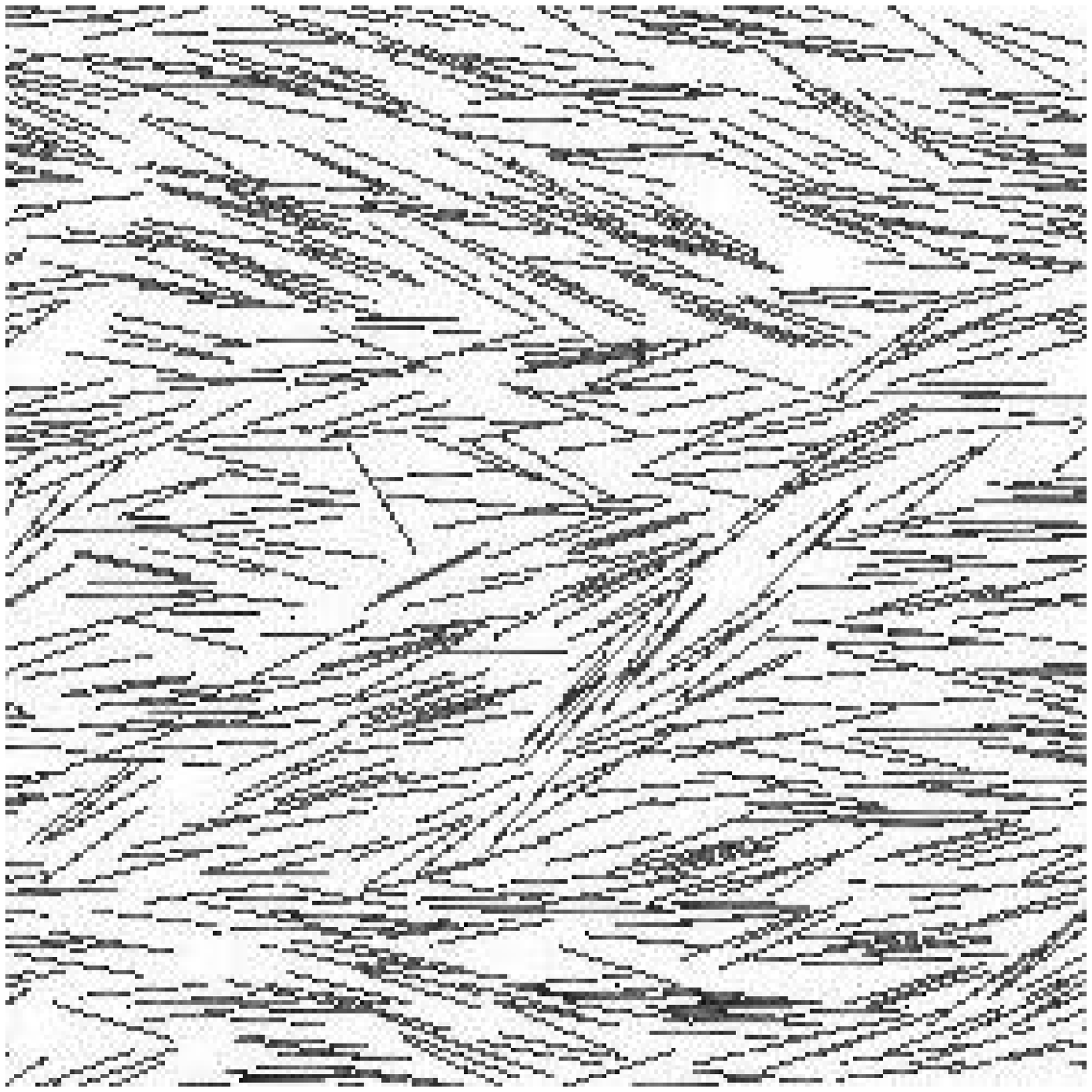}}
  \subfigure[]{\includegraphics[scale=.45]{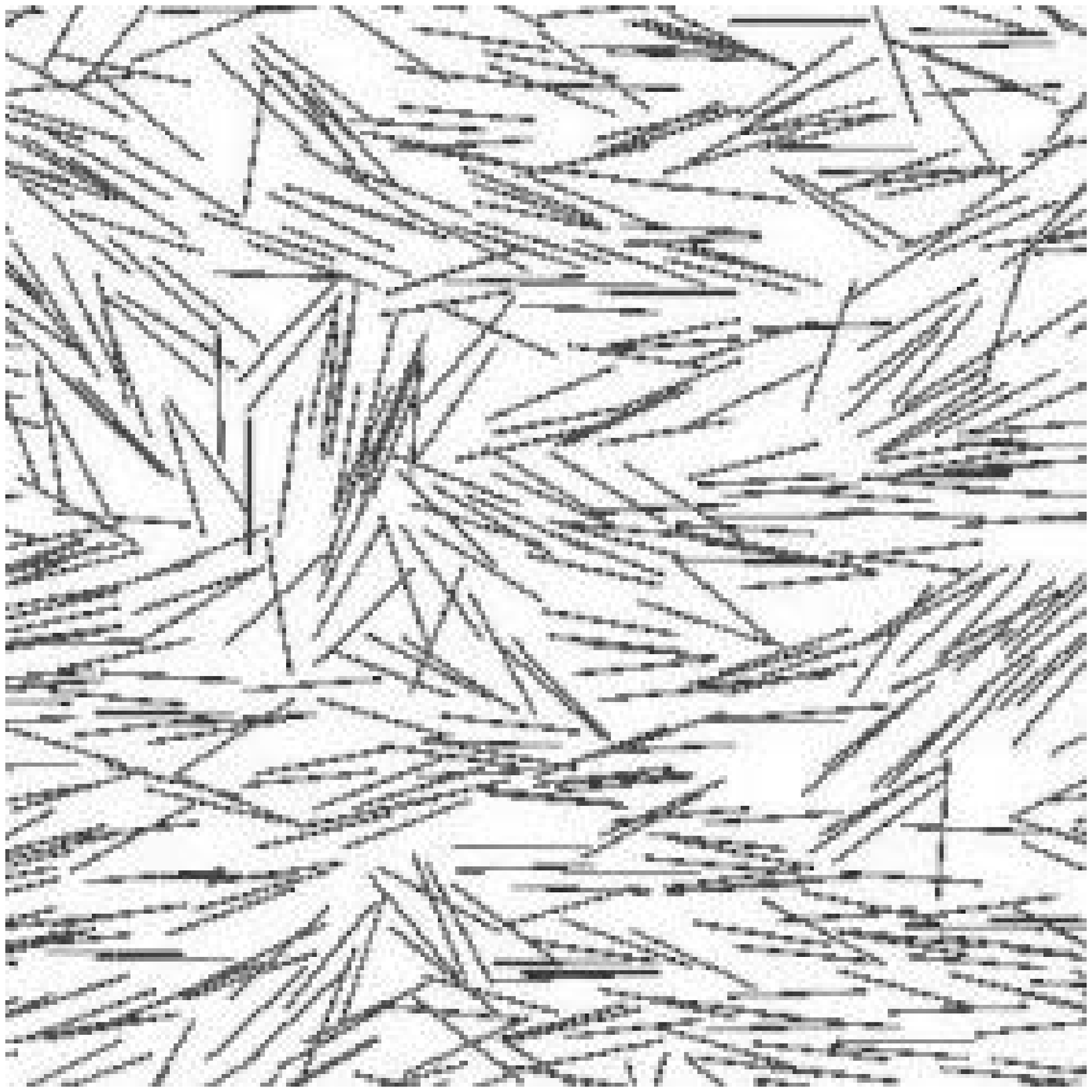}}
  \caption{Typical snapshots of configurations of two dimensional
    nematic-like phases (details). The pictures are projections of the
    configurations on the xy-plane. (a) 3074 particles in a slit with
    \( h=1\), \(L_{xy}=6000D\). The defects are noticeable in the
    picture. (b) 1150 particles in a slit with \(h=3\),
    \(L_{xy}=4000D\). Along with the defects, overlaps between
    projections of particles due to the transverse degrees of freedom
    are visible.}
  \label{fig:snap2D}
\end{figure}

From the above data, we can conclude that we are observing the same
phase transition as in the two dimensional system of needles,
described by the Frank elastic free energy,
\begin{displaymath}
  F = \frac{1}{2} K \int \nabla \theta({\bf r}) d{\bf r}.
\end{displaymath}
Therefore we can set the transition density \(C_{\textrm{trans}}^{KS}
\) when the exponent of \(g_2\) crosses \(1/4\)~\cite{frenkel}.  Going
to the limiting case \(h=1\), we can recover the fully two-dimensional
system and compare the transition density with the one given
in~\cite{frenkel}. We measure a \(C_{\textrm{trans}}^{KS}(h=1) = 6.5
\) which we consider in good agreement with the 7.5 to 8 of Frenkel
and Eppenga, considering the longer relaxation times we allow
(\(5\cdot 10^5\) sweeps for up to 2600 particles). This value can also
be compared with the \(5.3\) found in \cite{bates} for
disco-rectangles with \(L/D=15\).  The most interesting question,
though, is how the transition density changes as a small lateral
dimension is added. By a naive argument comparing the
three-dimensional density \(C\) with the two dimensional one \( C_{2d}
= \frac{N}{L_xL_y}(L+D)^2 \) one would expect to see a transition
density which goes like \(1/h\) so that the ``effective''
two-dimensional density is kept fixed.  However, according to our
data, shown in figure \ref{fig:KShdep}, this decay is slower than this
power law. We can try to understand this by a qualitative argument
based on the fact that this two dimensional transition is mediated by
defects. Adding a small transverse dimension, the liberty of effective
overlaps given by the third degree of freedom makes it harder to form
a disclination, so that a higher density than just \(h\) times the
two-dimensional one is required to achieve the transition. This fact
is illustrated in figure \ref{fig:snap2D}, where we show two typical
xy-projections of the nematic states for \(h = 1\) and \(h = 3\). The
configurations are qualitatively similar, but in the case \(h = 3\)
many effective overlaps are noticeable.

\subsection*{Broadly polydisperse rods}
\label{sec:exp}


We performed similar simulations for a system of polydisperse rods
with an exponential length distribution.  Our main motivation for
doing this was to predict the experimental behavior of microtubules
when they are confined in a region with a thickness that is very small
compared to their (mean) length and very large compared to their
diameter. The sample for this experiment consists of a microscope
coverslip patterned with \(1 \mu m\) thick photoresist used to
maintain a constant separation with the microscope slide (details will
be published elsewhere) \cite{marcodafa}. In the experiment
microtubules are broadly polydisperse, and typically show an
exponential distribution in length with an average aspect ratio of a
few hundreds.

\begin{figure}[!htb]
  \centering
  \includegraphics[scale=.45]{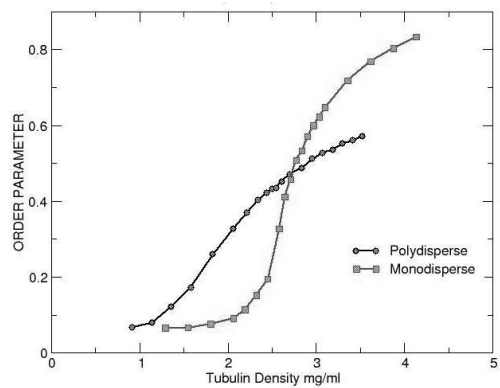}
  \caption{Apparent transition predicted by the simulation in the
    exponentially polydisperse case for a fixed size \(L_{xy} = 1200
    D\) (30 \(\mu m\)) compared to the monodisperse system with the
    same mean aspect ratio (320). This size is comparable with that of
    a field of view of the microscope. In the x-axis we plot the
    tubulin density estimated from the total length of the rods in the
    simulation box.}
  \label{fig:monopoly}
\end{figure}

In the polydisperse case, the apparent transition at fixed size is
broader than the one observed for the monodisperse system (figure
\ref{fig:monopoly}) and its onset is at lower densities, due to the
effect of long rods. Nevertheless, the qualitative behavior of the
system is the same, with size effects and power-law decay in the
correlation function.
By measuring the total length of the rods in one box, knowing that the
length of a tubulin dimer is 8 nm and that each polymer is composed of
13 protofilaments, one can predict the concentration of tubulin needed
to have the ordering transition at a fixed size. The size of the
system can be set as the field of view of the microscope, which is
effectively the scale at which the experimental system is observed.
With this identification, the apparent transition density, expressed
in terms of tubulin concentration, falls in the experimentally
accessible region of 2 to 5 mg/ml of tubulin.

\begin{figure}[!htb]
  \centering
  \includegraphics[scale=.95]{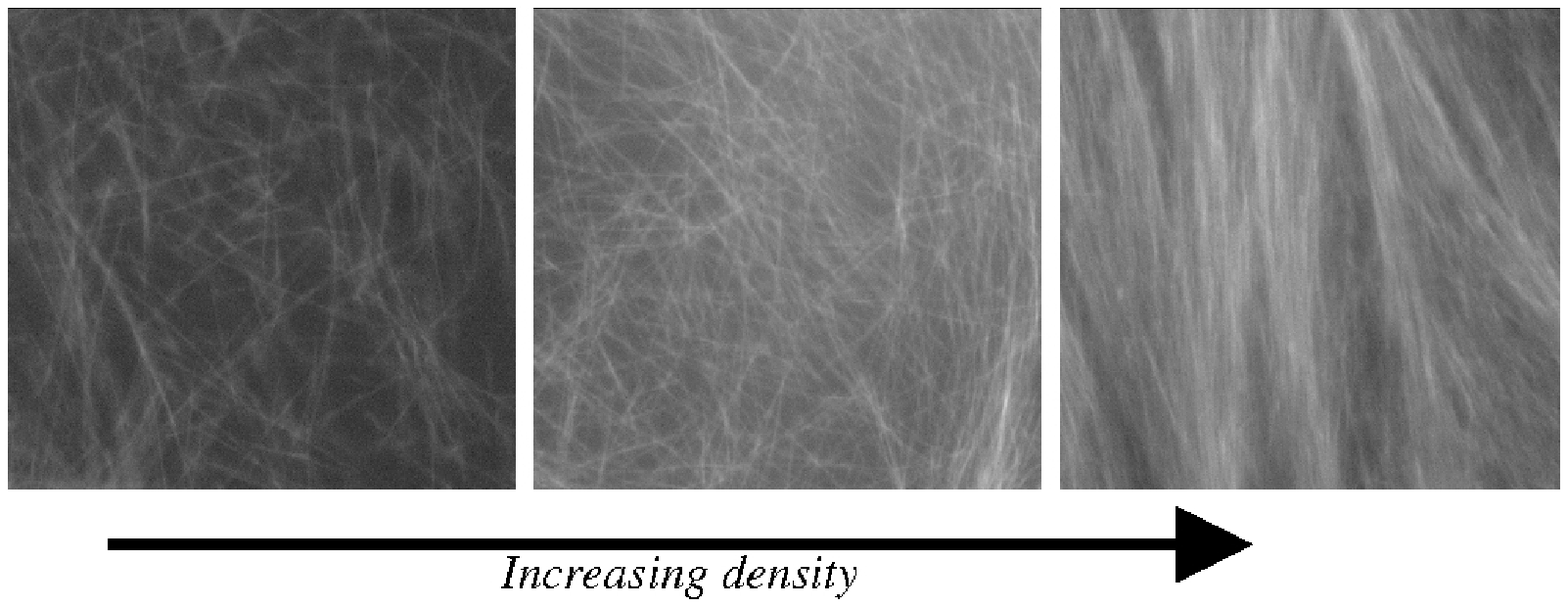}
  \caption{Snapshots from the experiment. The microtubules, confined
    in 1 \(\mu m\) thick microscope slides, are imaged with fluorescence
    microscopy. The concentration range of polymerized protein is 2-4
    mg/ml. The rods are observed to align for increasing densities on
    scales comparable to the field of view of the microscope (30-100
    \(\mu m\)). On larger scales, patches of aligned rods with
    different orientation are noticeable. }
  \label{fig:exper}
\end{figure}

These results are qualitatively consistent with those observed
experimentally (see figure \ref{fig:exper} as an illustration).  Also,
in the experiment, microtubules are observed to order in ``patches''
of tens of microns in size, which is consistent with the absence of
true long range order predicted by the simulation.  Quantitative work
on the experimental system is still in progress~\cite{marcodafa}.

\section{Conclusions and perspectives}
\label{sec:concl}

We presented a Monte Carlo simulation of hard spherocylinders confined
between two parallel walls at a small distance. This computation was
intended to test the behavior of the nematic transition in a
quasi-two-dimensional system.  The main result is that a transition to
long-range-order is absent in the thermodynamic limit which is
substituted by a Kosterlitz-Thouless-like transition revealed by
the algebraic decay of the orientational correlation function.  This
is in qualitative agreement with what is observed experimentally for a
broadly polydisperse system of rod-like biological polymers
(microtubules) and has some interesting general consequences.  First
of all, it indicates that a system with some small transverse third
dimension will maintain the same qualitative behavior as a true
two-dimensional one.  Secondly, the slower decrease of the
Kosterlitz-Thouless transition with increasing plate separations for
the 2D-like isotropic-nematic transition indicates that there is an
important role played by the transverse degree of freedom in the
elimination of disclinations.  Possible future work includes a
quantitative exploration of the experimental system. It is also
interesting to investigate in more detail the crossover between two-
and three-dimensional behavior, exploring slit thicknesses that are
comparable with the length of the rods. This kind of work would
require using spherocylinders with a lower aspect ratio, to keep the
number of particles, and therefore the computational cost, reasonably
low. One possibility that we can hypothesize is that the
Kosterlitz-Thouless transition in the strictly two-dimensional and
quasi-two-dimensional system ends in the previously observed
wall-induced uniaxial to biaxial transition~\cite{MarjoleinEPL} at
larger plate separation and is therefore not connected to capillary
nematization (a transition from a biaxial to a condensed nematic
phase), which terminates in a critical point at a fixed plate
separations~\cite{ReneJCP}.

\begin{acknowledgments}
  We would like to thank Bela Mulder and Daan Frenkel for help and
  discussion.  This work is part of the research program of the
  ``Stichting voor Fundamenteel Onderzoek der Materie (FOM)'', which
  is financially supported by the ``Nederlandse organisatie voor
  Wetenschappelijk Onderzoek (NWO)''.
\end{acknowledgments}


\begin{thebibliography}{99}


\bibitem{marcel} F. Gittes, \emph{et.al.}, J. Cell Biol. \bf{120}, 923 (1993).


\bibitem{frenkel}  D.Frenkel and  R.Eppenga, Phys. Rev. A \bf{31} (3),
  1776 (1985).

\bibitem{bates}  M.A.Bates and D. Frenkel, J. Chem. Phys. \bf{112} (22),
  10034 (2000).


\bibitem{koth} J.M. Kosterlitz and D. Thouless. J. Phys. C \bf{6} 1181 (1973).


\bibitem{schmidt_def}  J.Dzubiella,  M.Schmidt,  H.L\"owen,
  Phys. Rev. E, \bf{62} 5081 (2000).


\bibitem{schmidt_sph} M. Schmidt and L\"owen H., Phys. Rev. Lett., \bf{76}
  4552 (1996).

\bibitem{poniewirski} A. Poniewirski,  Phys. Rev, E \bf{47} 3396 (1993).


\bibitem{MarjoleinEPL} R. Van Roij, M. Dijkstra, R. Evans,
Europhys. Lett. \bf{49} (3), 350 (2000).


\bibitem{chen}  Z.Y. Chen and  S.M.Cui, Phys. Rev. E \bf{52} 3876 (1995).

\bibitem{ReneJCP} R. van Roij, M. Dijkstra, and R. Evans, J. Chem. Phys. 
{\bf 113}, 7689 (2000). 

\bibitem{MarjoleinPRE} M. Dijkstra, R. van Roij, R. Evans, Phys. Rev.
  E \bf{63} (5) 051703 (2001).
  

\bibitem{marcodafa} M. Cosentino Lagomarsino , M.Dogterom,
 to be published.

\end{thebibliography}
\end{document}